\documentclass[fleqn,usenatbib]{mnras}

\usepackage{newtxtext,newtxmath}

\usepackage[T1]{fontenc}
\usepackage{ae,aecompl}

\usepackage{graphicx}	
\usepackage{amsmath}	
\usepackage{amssymb}	

\title[Extreme outflows in radio-loud NLS1s]{Extreme gaseous outflows in radio-loud narrow-line Seyfert 1 galaxies}

\author[S. Komossa et al.]{
S. Komossa,$^{1}$\thanks{E-mail: astrokomossa@gmx.de}
D.W. Xu,$^{2,3}$
and A.Y. Wagner$^{4,5}$
\\
$^{1}$Max-Planck-Institut f\"ur Radioastronomie, Auf dem H{\"u}gel 69, 53121 Bonn, Germany\\
$^{2}$Key Laboratory of Space Astronomy and Technology, National Astronomical Observatories, CAS,
Beijing 100012, China\\
$^{3}$University of the Chinese Academy of Sciences, 19A Yuquan Road, Shijingshan District, Beijing 100049, China \\
$^{4}$Center for Computational Sciences, University of Tsukuba, 1-1-1 Tennodai, Tsukuba, Ibaraki 305-8577 Japan\\ 
$^{5}$Institut d'Astrophysique de Paris, 98 bis bd Arago, F-75014 Paris, France  } 

\date{Accepted 2018 April 7. Received 2018 April 1; in original form 2018 February 15}

\pubyear{2015}

\begin{document}
\label{firstpage}
\pagerange{\pageref{firstpage}--\pageref{lastpage}}
\maketitle

\begin{abstract}
We present four radio-loud NLS1 galaxies
with extreme emission-line shifts, indicating radial outflow velocities 
of the ionized gas of up to 2450 km/s, above the escape velocity of the host galaxies. The 
forbidden lines show strong broadening,  
up to 2270 km/s.   
An ionization stratification (higher line shift at higher ionization
potential) 
implies that we see a large-scale outflow rather than single, 
localized jet-cloud interactions. 
Similarly, the paucity of zero-velocity [OIII]$\lambda$5007 emitting gas implies the absence of a 
second narrow-line region (NLR) component at rest, and therefore  
a large part of the high-ionization NLR is affected
by the outflow. 
Given the radio loudness of these NLS1 galaxies, the observations are consistent with
a pole on view onto their central engines, so that the effects of polar outflows are maximized. 
In addition, a very efficient driving mechanism is required, to reach the high 
observed velocities.  
We explore implications from recent hydrodynamic simulations of the interaction between
fast winds or jets with the large-scale NLR. Overall, the best agreement with observations
(and especially the high outflow speeds of the [NeV] emitting gas) 
can be reached if the NLS1 galaxies are relatively young sources with lifetimes not much
exceeding 1 Myr.   
These systems represent sites of strong feedback at NLR scales at work, well below redshift one.  
\end{abstract}

\begin{keywords}
galaxies: active -- galaxies: evolution -- 
galaxies: individual (SDSSJ130522.75+511640.3, SDSSJ144318.56+472556.7, SDSSJ150506.48+032630.8,
SDSSJ163401.94+480940.2) -- galaxies: jets -- 
galaxies: Seyfert -- quasars: emission lines
\end{keywords}



\section{Introduction}

Powerful gaseous outflows in Active Galactic Nuclei
(AGN) deposit mass, energy and metals in the interstellar
medium of the host galaxy or on even larger scales
(e.g., Colbert et al. 1996, Churazov et al. 2001, Moll
et al. 2007, Hopkins et al. 2016). They therefore shape the structure and
composition of the AGN environment, and may play an
important role in unified models (e.g., Elvis 2006).

The most powerful of these outflows can significantly affect 
the co-evolution of galaxies and black
holes by feedback processes (e.g., Fabian 1999, Wyithe
\& Loeb 2003), especially by regulating star formation,
and possibly clearing the host galaxy of large fractions
of its ISM (e.g., di Matteo et al. 2005, Springel et al.
2005, Hopkins \& Elvis 2010, Zubovas \& King 2012, Hopkins et al. 2016).

\begin{table*}
\centering
\caption{Galaxy radio properties. Columns from left to right: (1) SDSS optical
J2000 coordinates in RA (h,m,s) and DEC (d,m,s), (2) galaxy name used in this work, (3) SDSS redshift,
(4) radio-loudness index $R$ (Kellermann et al. 1989) defined
as $f_{\nu}$(1.4 GHz)/$f_{\nu}$(4400\AA) from Yuan et al. (2008), (5)--(8) radio flux densities in mJy at
151 MHz, 1.4 (FIRST/NVSS), 4.85 and 22 GHz collected from the literature (Hales et al. 2007, Helfand et al. 2015,
Condon et al. 1998, Gregory \& Condon 1991, Doi et al. 2016),
(9) spectral slope $\alpha_r$ between 1.4 GHz (FIRST) and 4.85 GHz, based on non-simultaneous data.}
\label{tab:Table1}
   \begin{tabular}{ccccccccc} 
   \hline
   coordinates & abbreviated name & $z$ & log $R$ & $f_{\rm 0.15GHz}$ & $f_{\rm 1.4GHz}$ & $f_{\rm 4.85GHz}$ & $f_{\rm 22GHz}$ & $\alpha_r$  \\
   (1)  & (2) & (3) & (4) & (5) & (6) & (7) & (8) & (9)  \\ 
   \hline
   13:05:22.75  +51:16:40.3 & SDSSJ1305+5116 & 0.785 & 2.34 & 320 & 87/87   & 46  & $<$9  & $-$0.51    \\
   14:43:18.56  +47:25:56.7 & SDSSJ1443+4725 & 0.703 & 3.07 & 359 & 171/166 & 59  & $<$16 & $-$0.86    \\
   15:05:06.48  +03:26:30.8 & SDSSJ1505+0326 & 0.409 & 3.19 & 120 & 380/395 & 991 & 697   &   0.77    \\
   16:34:01.94  +48:09:40.2 & SDSSJ1634+4809 & 0.495 & 2.31 & ... &   8/14  & ... & ...   & ... \\ 
		\hline
	\end{tabular}
\end{table*}

There is ample observational evidence for mild winds
and outflows in AGN (e.g., Crenshaw \& Kraemer 2007, Fabian 2012).
In the optical and IR regime, they manifest as emission-line
shifts and line asymmetries. 
Narrow emission lines in nearby AGN typically imply
outflow velocities of less than a few 100 km s$^{-1}$ in the high-ionization gas, 
even though much higher values have 
occasionally been reported, especially seen in `blue wings'
of the [OIII]$\lambda$5007  emission line (e.g., Capetti et al. 1999, Holt et al. 2003, 
Das et al. 2005, Das et al. 2006, Kraemer et al. 2009, 
Holt et al. 2008, Mazzalay et al. 2010, Nesvadba et al. 2011,
Villar-Martin et al. 2011, Jin et al. 2012, 
Bae \& Woo 2014, Komossa et al. 2015, Shen 2016, Woo et al. 2016, Perna
et al. 2017, Wang et al. 2018). 
Most of these do not affect the whole narrow-line region (NLR), but occur in 
localized regions, sometimes though not always spatially
coincident with radio jets. Longslit and integral field
spectroscopy have been powerful tools in uncovering them.
In other instances, the outflows are widely extended, and/or involve
the whole [OIII] emission (e.g., Zamanov et al. 2002, 
Marziani et al. 2003, Boroson 2005, Komossa et al.
2008, Zhang et al. 2011, Bae \& Woo 2014, Harrison et al. 2014, Cracco et al. 2016, 
Berton et al. 2016, Marziani et al. 2016, Zakamska et al. 2016, Rupke et al. 2017). 
Galaxies which show a systematic shift of their whole [OIII] emission line
with respect to H$\beta$ have been termed `blue outliers' (Zamanov et al. 2002).
All galaxies of our study are blue outliers.

The driving force of these outflows has remained unclear. Accretion-disc winds, 
radiation pressure acting on dusty gas, entrainment of gas in radio plasma, and
variants thereof, have all been considered (e.g., Murray et al. 1995,
Binette 1998, Proga et al. 2000, Saxton et al. 2005,
Proga et al. 2008, Wagner et al. 2012, 2013, Thompson et al. 2015, 
Bieri et al. 2017, Costa et al. 2018, Cielo et al. 2018).

There are indications, that winds are especially strong
in NLS1 galaxies, where accretion near the Eddington
limit likely triggers strong, radiation-pressure driven out-
flows (e.g., Boroson 2002, Grupe 2004, Grupe et al. 2010,
Xu et al. 2012). 
NLS1 galaxies are a subclass of AGN with extreme multi-wavelength
properties. Common definition criteria are small widths of their broad Balmer lines of
FWHM(H$\beta$) $<$ 2000 km s$^{-1}$, faint emission of [OIII]/H$\beta < 3$,
and strong FeII emission complexes (Osterbrock \& Pogge 1985, Goodrich 1989,
Veron-Cetty et al. 2001).
About 16\% of all NLS1 galaxies are blue outliers and exhibit strong
kinematic shifts of their [OIII] emission-line cores 
($v > 150$ km s$^{-1}$, Komossa et al. 2008). 

While NLS1 galaxies are on average more radio-quiet
than broad-line AGN (Komossa et al. 2006), a small
fraction of them is beamed and radio-loud (Komossa et al. 2006, 
Yuan et al. 2008),
highly variable at radio frequencies (Abdo et al. 2009c, L{\"a}hteenm{\"a}ki et al. 2017), and 
detected at $\gamma$-rays (Abdo et al. 2009a,b, 
Foschini 2011, D'Ammando et al. 2012, 2015, Yao et al. 2015, Paliya et al. 2018, Yang et al. 2018).

If radio loudness of NLS1 galaxies is generally due to beaming, 
we expect a pole on view
onto these galaxies, and so the effects of (polar) outflows are maximized.
We have analyzed four radio-loud NLS1 galaxies of
the sample of Yuan et al. (2008) which were initially noted 
for their shifted [OIII] emission lines, placing them in the class of blue outliers.  
We show that all four galaxies of our study exhibit extreme line shifts, among
the highest measured so far. They therefore provide us with powerful
tools of understanding extreme gaseous outflows in NLS1 galaxies; their drivers, possible
association with radio jets, and their impact on the environment.
 This is the first dedicated
study of highly-ionized gas outflows in the radio-loudest NLS1 galaxies 
including emission lines other than the [OIII]$\lambda$5007--H$\beta$ complex.
First results were reported by Komossa et al. (2016). 
 
This paper is organized as follows. In Sect. 2 we
describe the methods of data analysis and present the results. 
Notes on individual galaxies are given in Sect. 3. 
We then confront the large measured emission-line shifts and
widths with different outflow mechanisms 
 (Sect. 4). Conclusions are provided in Section 5.
We use a cosmology (Wright 2006) with 
$H_{\rm 0}$=70 km\,s$^{-1}$\,Mpc$^{-1}$, $\Omega_{\rm M}$=0.3
and $\Omega_{\rm \Lambda}$=0.7 throughout this paper. 

Finally, we would like to emphasize, that [OIII] line shifts come in 
two types, and we keep this distinction throughout this paper: The [OIII] emission 
line can typically be fit by {\em two} Gaussian components;
one strong prime component which we refer to as narrow ``core component'' of the emission line,
and a second, fainter and broader component which is often shifted toward the
blue w.r.t. the core component, 
and is commonly referred to as blue wing.
Many studies of [OIII] outflows are based on just these blue wings, while in blue outliers
(including those discussed here), the {\em whole core profile} of [OIII] is shifted.  
Blue wings are additionally
present in our sources, and show extreme shifts and widths.

\section{Data analysis and results}

\subsection{Data preparation}

\begin{figure*}
\includegraphics[width=14cm]{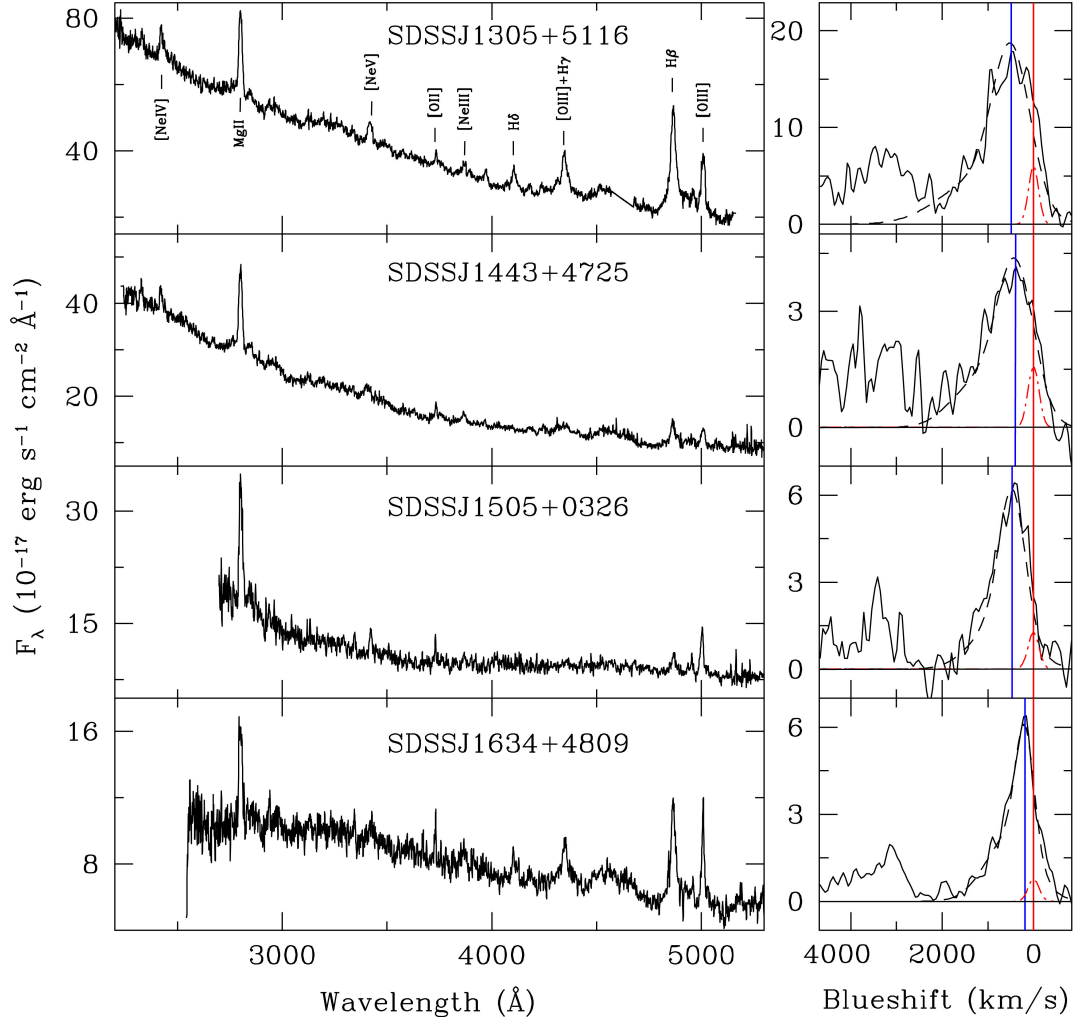}
\caption{Left panel: SDSS spectra of the four radio-loud NLS1 galaxies (rest-frame wavelength based on SDSS redshift).
Right panel: zoom on the [OIII]$\lambda\lambda$4959,5007
emission-line doublet of the four galaxies, after
subtraction of FeII emission and continuum. The [OIII] line blueshift is measured with respect to [OII].
The red dotted line marks the maximum contribution of a Gaussian with zero blueshift. The solid blue line
marks the line peak location based on Gaussian fitting.
See the electronic edition of the Journal for a colour version of this figure.
 }
 \label{o3profiles}
 \end{figure*}

The four galaxies (Tab. 1) have been observed in the course of the
Sloan Digital Sky Survey (SDSS; York et al. 2000,
Zhou et al. 2006, Schneider et al. 2007, Shen et al. 2011), and we have retrieved the
spectra from data release DR7 (Abazajian et al. 2009) for analysis.
The data preparation and spectral analysis has been performed in a standard way,
similar to Xu et al. (2007).  The spectra were corrected for Galactic extinction, according to
Schlegel et al. (1998).  All spectra show strong emission complexes from FeII
in the optical and UV, and therefore FeII templates were prepared and
subtracted before further emission-line analysis. In the optical,
the FeII template of Veron-Cetty et al. (2004) was used, while in the UV
we employed the template of Bruhweiler \& Verner (2008). It was assumed
that FeII has the same profile as the broad component of H$\beta$.
When we report measurements of optical FeII strength, FeII\,$\lambda$4570, this is
the integrated flux of the FeII emission complex between
the rest wavelengths 4434\AA\ and 4684\AA. R4570 
corresponds to the ratio FeII\,$\lambda$4570/H$\beta$.
Some basic properties of the four galaxies are listed in Tab. 1, including
source redshift, and radio properties collected from the literature.

\begin{figure}
\includegraphics[width=\columnwidth]{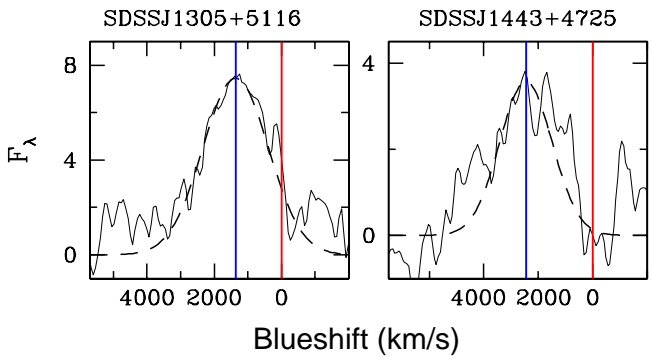}
\caption{Observed [NeV]$\lambda$3426 line profile of the two galaxies with the highest outflow velocities,
and fitted Gaussian profile. 
}
\label{nev} 
\end{figure}

\subsection{Emission-line fits} 

The spectra show emission lines from H$\beta$, [OII]$\lambda$3727,
[OIII]$\lambda\lambda$5007,4959, [NeIII]$\lambda$3869, [NeV]$\lambda$3426, 
MgII\,$\lambda$2798, and CII]$\lambda$2325.
SDSSJ1305+5116 and SDSSJ1443+4725 further show a line near 2424\AA; 
likely due to [NeIV] with a possible contribution from FeIII (Sect. 2.8). 
Emission line fits were carried out with the IRAF package SPECFIT (Kriss 1994).
Emission lines were fit with Gaussian profiles, in a way similar to Komossa et al. 2008 (K08
hereafter),
which also allows easy comparison with that sample of NLS1 galaxies and
blue outliers in particular.
The one difference with the blue outliers of K08 is the higher redshifts of the four galaxies
analyzed here, and so the [SII]$\lambda$$\lambda$6716,6731 lines used previously for reference 
when measuring kinematic shifts, are no longer observable. Instead, we have therefore measured
line shifts with respect to [OII]$\lambda$3727.

Emission-line shifts, FWHMs,
and fluxes were measured. Results are reported in Tab. 2. 
All FWHMs given in this paper have been
corrected for instrumental broadening. 
Most emission lines are well represented by a single Gaussian profile. 
Exceptions are the brightest emission lines, [OIII], H$\beta$ and MgII. 
H$\beta$ was decomposed into three components: a narrow      
component (H$\beta_{\rm n}$), and two broad components.
No physical meaning is ascribed to the two separate
broad components; they merely serve as a mathematical
description (a Lorentzian often is an alternative to describe
such profiles; e.g.,  Veron-Cetty et al. 2001, Sulentic et al. 2002).
The final width of the broad-line emission, H$\beta_{\rm b}$, reported in Tab. 2
and throughout this paper, 
is then determined
as the FWHM of the sum of those two Gaussians.

The total [OIII] emission-line profile, [OIII]$_{\rm totl}$,
was decomposed into two Gaussian components,
a narrow core ([OIII]$_{\rm c}$) and a broad `wing' component ([OIII]$_{\rm w}$).
Measurements of the FWHM and blueshift of [OIII] reported in this work
refer to the core of the emission line, unless noted otherwise.
MgII was fit with two Gaussian components, a narrow and a broad component.
Results are reported in Tab. 2.
The optical spectra of all four galaxies are shown in Fig. 1, while a zoom on the [NeV] emission
of the two galaxies with the highest outflow velocities is given in Fig. 2. 

\begin{table}
\centering
\caption{Emission-line measurements. Columns from left to right: (1) Emission line.
(2) Emission-line velocity (outflow: positive sign),
measured relative to [OII], in km s$^{-1}$.  (3) Measured line flux, 
in 10$^{-15}$ erg cm$^{-2}$ s$^{-1}$.
(4) Line width (FWHM), corrected for instrumental broadening, in km s$^{-1}$.
$^{a}$Line identification uncertain. See Sect. 2.8 for details.}
\label{tab:Table2}
   \begin{tabular}{lccc} 
   \hline
galaxy name / line & $\Delta v$ & $f$ &  FWHM \\
 (1) & (2) & (3) & (4) \\ 
\hline
SDSSJ1305+5116 &   &   &  \\
\hline
~MgII$_n$         & 460  & 0.17 & 500  \\
~MgII$_b$         & 400  & 9.37 & 2170 \\
~CII]             & -50  & 1.23 &  1220 \\
~[NeIV]$^{(b)}$   &  660 & 2.66 & 1710  \\
~[NeV]            & 1360 & 3.71 & 2300 \\
~[OII]            & 0    & 0.70   & 650  \\
~[NeIII]          & 640  & 1.19 & 1190 \\
~[OIII]$_c$       & 480  & 4.85 & 950\\
~[OIII]$_w$       & 1230 & 2.71 & 1950\\
~H$\beta$$_n$     & 150  & 0.49 & 500\\
~H$\beta$$_b$     & 270  & 22.6 & 1990\\
~R4570            &  & 0.62  & \\
\hline
SDSSJ1443+4725 & &  &  \\
\hline
~MgII$_n$        & 230  & 0.10 &  200 \\
~MgII$_b$        & 250  & 7.01 & 2300 \\
~CII]            & 220  & 0.72 &  950 \\
~[NeIV]$^{a}$  & 750 & 0.99 & 1100 \\
~[NeV]           & 2450 & 1.67 &  2270 \\
~[OII]           & 0    & 0.35 &  490 \\
~[NeIII]         & 760  & 0.62 & 1140 \\
~[OIII]$_c$      & 400  & 1.18 & 940\\
~[OIII]$_w$      & 1280 & 0.42 & 1300 \\
~H$\beta$$_n$    & 420  & 0.20 & 290 \\
~H$\beta$$_b$    & 340  & 2.79 & 1700\\
~R4570           &      & 2.2 & \\
\hline
SDSSJ1505+0326 & & &  \\
\hline
~MgII$_n$        & 170 & 0.05 & 260  \\
~MgII$_b$        & 170 & 3.73 & 1500 \\
~[NeV]           & 680 & 0.68 & 1070 \\
~[OII]           & 0   & 0.23 &  270 \\
~[OIII]$_c$      & 460 & 0.78 & 640\\
~[OIII]$_w$      & 720 & 0.57 & 1390\\
~H$\beta$$_n$    & -80 & 0.05 & 250\\
~H$\beta$$_b$    & -110 & 0.84 & 1440 \\
~R4570           &      & 1.63 & \\
\hline
SDSSJ1634+4809 & &  &  \\
\hline
~MgII$_n$        & 90  & 0.05 & 400  \\
~MgII$_b$        & 90  & 1.54 & 1840 \\
~[NeV]           & 520 & 0.20 & 1270 \\
~[OII]           & 0   & 0.20 & 400  \\
~[NeIII]         & 460 & 0.32 & 1140 \\
~[OIII]$_c$      & 190 & 0.55 & 450 \\
~[OIII]$_w$      & 520 & 0.72 & 1220\\
~H$\beta$$_n$    &  80 & 0.21 & 370 \\
~H$\beta$$_b$    &  10 & 3.61 & 1770 \\
~R4570           &  & 1.02 & \\ 
\hline
	\end{tabular}
\end{table}

\subsection{Limits on the presence of a zero-velocity [OIII] component, and of highly blueshifted H$\beta$} 

We have used the spectrum of SDSSJ1305+5116, the brightest of the four sources,
for some further measurements. 
First, we have checked for the presence of an H$\beta$ ``counterpart'' to 
the blueshifted [OIII] core component. We did so by refitting the H$\beta$ profile
with an extra Gaussian, its FWHM and peak shift fixed to that of [OIII]. The
two remaining broad (and the narrow) Gaussians were left free to vary.
This procedure allows us to estimate the maximum emission from any highly
blueshifted H$\beta$, if any, consistent with the spectrum. We find that the
presence of blueshifted H$\beta$ is consistent with the spectrum of SDSSJ1305+5116,
with a maximum contribution of 15\% in flux.  

Second, in order to see, how much zero-velocity [OIII] emitting gas there is, we have
also fit an extra Gaussian to [OIII] fixed at a narrow width
of only FWHM=230 km s$^{-1}$ (the average width of low-ionization 
emission lines in the sample of K08), and at the same redshift as [OII].
We find little if any contribution of zero-velocity
[OIII], with an upper limit of typically 1--5\% of the total flux in [OIII]. 

Finally, the broad Balmer-line component, H$\beta_{\rm b}$,  
was inspected for a possible blueshift. 
In our fitting procedure, the broad part of the emission line was approximated
by two Gaussian components. On average, these do not show
significant blue or redshifts beyond $\sim$300 km s$^{-1}$, and always remain 
below the high shifts seen in the high-ionization emission lines.  
Therefore, the bulk of the H$\beta$-emitting BLR does not participate in 
the high-velocity outflow. 

\subsection{Measurement uncertainties}

In order to see, how much the measurement of the narrow emission line parameters 
are affected by emission-line decomposition and by the way
the continuum is modelled, 
we have performed various tests, including different
ways of measuring line widths and peak shifts, 
and fitting only a fraction of the full profile.
We find that uncertainties in 
line widths and shifts are typically less than 10--20\%. 

The NLS1 galaxies of our study are also included in a number of large SDSS NLS1/quasar
catalogues, which also list H$\beta$ line fits. We have compared our fit results
with those previous reports and find overall good agreement, given the fact that 
(1) those previous studies were standardized line fittings for very large samples,
and (2) different line decompositions were employed including (a) Lorentzian fits
with or without adding a narrow-line component, or (b) multiple Gaussian fits (one to three
components) with or without including a NLR component. For instance, for SDSSJ1305+5116,
Zhou et al. (2006) reported FWHM(H$\beta$)= 1952 km s$^{-1}$, which compares to 
1963 km s$^{-1}$ of Rakshit et al. (2017) and 1990 km s$^{-1}$ (our Tab. 3).  
Berton et al. (2016) analyzed the [OIII] profile
of a sample of NLS1 galaxies.  
While they determined 
FWHM([OIII]$_{\rm{c}}$) = 1012 km s$^{-1}$ 
for SDSSJ1305+5116, we derived 950 km s$^{-1}$.  

\begin{figure}
\includegraphics[width=\columnwidth]{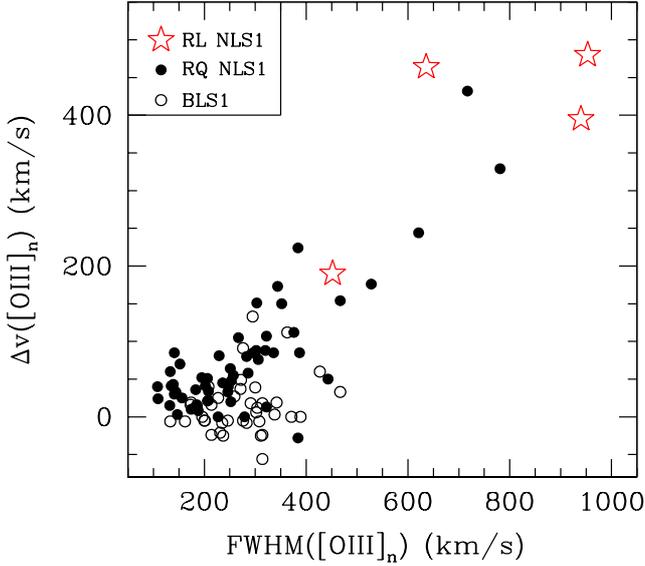}
\caption{Shift-width correlation of the core of [OIII]$\lambda$5007. For comparison,
the sample of BLS1 galaxies (open circles) and NLS1 galaxies (filled circles)
of Komossa et al. (2008) is plotted.
}
\label{corr}
\end{figure}

\subsection{NLS1 classification} 

We have always used Gaussians to fit H$\beta$. We confirm the
NLS1 classification (Zhou et al. 2006) of all four galaxies,
since always FWHM(H$\beta_{\rm b}$) $<$ 2000 km s$^{-1}$. However, it is interesting
to note that the {\em broadest} component of H$\beta$ (we recall that
we fit each H$\beta$ with two broad components, and the FWHM represents the sum
of those two) of SDSSJ1634+4809
is relatively broad, with FWHM(H$\beta_{\rm b2}$)=4960 km s$^{-1}$.  
This reflects
the fact, that AGN with narrow broad lines are often
well fit with Lorentzian profiles (e.g., Veron-Cetty 
et al. 2001, Sulentic et al. 2002). 

\subsection{Extreme velocities and line widths} 

Extreme emission-line blueshifts are present in all four spectra. 
In [OIII], SDSSJ1305+5116 and SDSSJ1505+0326 show the highest,
with FWHM([OIII]$_{\rm c}$) = 950 km s$^{-1}$ and $\Delta v$([OIII]$_{\rm c}$) = 480 km s$^{-1}$
(SDSSJ1305+5116) and FWHM([OIII]$_{\rm c}$) = 640 km s$^{-1}$ and 
$\Delta v$([OIII]$_{\rm c}$) = 460 km s$^{-1}$
(SDSSJ1505+0326). 
Blue wings in [OIII] are additionally present, and these show
even higher shifts and widths; reaching FWHM([OIII]$_{\rm w}$) = 1950 km s$^{-1}$ and
$\Delta v$([OIII]$_{\rm w}$) = 1230 km s$^{-1}$ in the case of SDSSJ1305+5116. 
 
High-ionization lines of [NeV] are identified in all spectra. 
The highest measured shifts are $\Delta v$([NeV]) = 2450 km s$^{-1}$
 (SDSSJ1443+4725) and $\Delta v$([NeV]) = 1360 km s$^{-1}$ (SDSSJ1305+5116). 

The [OIII] emission lines follow a width-shift correlation, in the sense that more
highly shifted lines are much broader (Fig. \ref{corr}), similar to the trend seen
in radio-quiet NLS1 galaxies and higher-redshift quasars. 
The same pattern is followed by [NeV]. Such a trend can be reproduced, if outflows
of higher velocity come with larger outflow-cone opening angles (Bae \& Woo 2016),
or with higher gas turbulence. Since in these cases, the emission-line width does
not reflect the host bulge potential, FWHM([OIII]) is not a good surrogate for stellar
velocity dispersion (Komossa \& Xu 2007), and therefore cannot be used for independent
BH mass estimates (Sect. 2.9) of these radio-loud NLS1 galaxies.

\subsection{A correlation of line shift with ionization potential}

\begin{figure}
\includegraphics[width=\columnwidth]{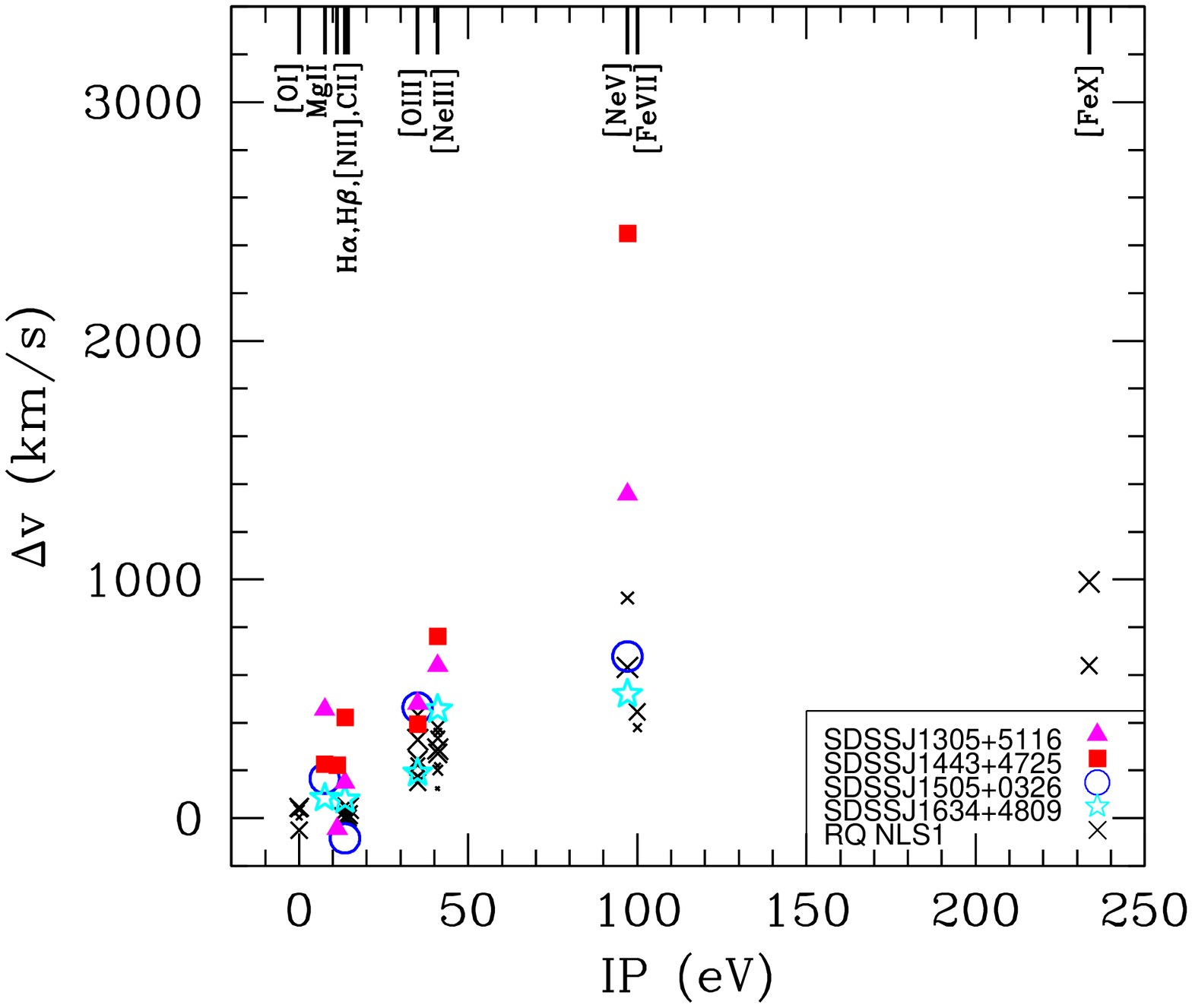}
 \caption{Radial velocity as a function of ionization potential IP.
The four galaxies of this study are
marked by large red and blue symbols. For comparison, we plot the nine blue outliers of
Komossa et al. (2008) with small black crosses.
See the electronic edition of the Journal for a colour version of this figure.
 }
 \label{ionstrat}
\vskip0.5cm

\includegraphics[width=\columnwidth]{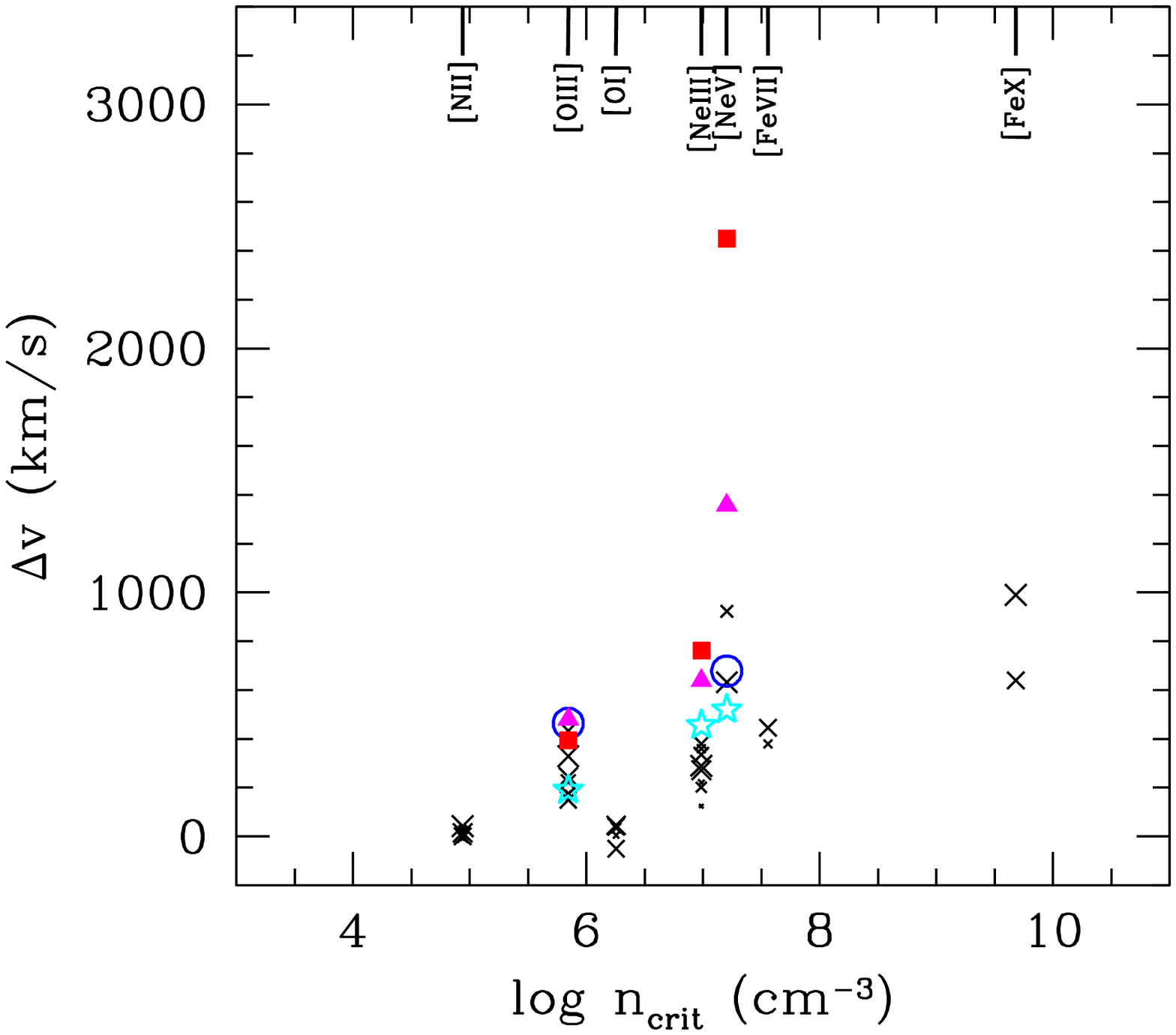}
 \caption{Radial velocity as a function of critical
density $n_{\rm crit}$.
The four galaxies of this study are
marked by large red and blue symbols. For comparison, we plot the nine blue outliers
of Komossa et al. (2008) with small black crosses.
See the electronic edition of the Journal for a colour version of this figure. }
 \label{ncrit}
\end{figure}

The higher-ionization lines show the higher blueshifts. We have therefore
plotted in Fig. {\ref{ionstrat}}  the dependence of radial velocity $\Delta v$ on ionization
potential IP of each respective ion, in comparison to our previous sample
of blue outliers (K08). Despite some scatter in the low-ionization
lines, and especially MgII, there is an overall trend of higher blueshift with
higher ionization potential{\footnote{note the tendency in several (but not all) nearby
Seyfert galaxies that high-ionization lines are typically more blueshifted than low-ionization
lines (e.g., Penston et al. 1984, De Robertis \& Osterbrock 1984,
Fig. 2 of Kraemer \& Crenshaw 2000, Fig. 17 of Mazzalay et al. 2010)}}. 

Fig. 5 shows the dependence on critical density. 
By chance, higher ionization potentials of the ions discussed here also
generally come with higher critical densities of the forbidden line transitions
observed from the respective ions. 
Therefore, a correlation between outflow
velocity and ionization potential generally
implies a correlation with critical density, and vice versa, raising
the question which of the two is the fundamental one;
density stratification or ionization stratification.  
An important exception to the above rule is [OI]$\lambda$6300.
Its critical density ($n_{\rm crit}=1.8\,10^{6}$ cm$^{-3}$) is relatively high,
while its ionization potential is zero. Unfortunately, [OI] is not observed 
in the four radio-loud NLS1 galaxies because of their high redshift. If the same
driver of the outflow is at work as in the radio-quiet NLS1 galaxies, then   
the deviation of [OI] (Fig. 5) suggests that
the trend with IP is the more fundamental one.

\subsection{Identification of the line near 2424\AA} 

The spectra of SDSSJ1305+5116 and SDSSJ1443+4725 show an emission line at a rest wavelength
of $\lambda_{\rm o}$ = 2422.1\AA~ and 2420.8\AA, respectively. 
This feature is near the location of [NeIV], which has been
previously identified in the SDSS spectra of AGN,
at a laboratory
wavelength of 2424\AA~(Tab. 2 of vanden Berk et al. 2001).  
If we do identify the observed feature in the spectra of
SDSSJ1305+5116 and SDSSJ1443+4725 with [NeIV], then outflow velocities of 
$\Delta v$(NeIV) = 660 km s$^{-1}$ and 750 km s$^{-1}$
(SDSSJ1305+5116 and SDSSJ1443+4725, respectively) are inferred. 
Given the possibility of a blend with FeIII transitions 
(vanden Berk et al. 2001), we do not include [NeIV] in the figures. 
We do list it in Tab. 2, and line parameters in that table have been inferred
under the assumption that we indeed see [NeIV].   

\subsection{Black hole masses and Eddington ratios}

Supermassive black hole (SMBH) masses were estimated based on the virial relation established for broad-line AGN,
using the line width of H$\beta_{\rm b}$. Since the optical continuum luminosity is likely
dominated by the jet emission, we make use of the H$\beta$ line luminosity instead 
(Vestergaard \& Peterson 2006):

\begin{equation} 
M_{\rm BH} = 10^{6.67}\bigg{(}\frac{L_{\rm H\beta}}{10^{42}\,{\rm erg\,s^{-1} }}\bigg{)}^{0.63} \bigg{(}\frac{\rm FWHM(H\beta)}{1000\, \rm{km\,s^{-1}}}\bigg{)}^2\, \rm{M_\odot}.
\end{equation} 

This gives SMBH masses in the range (6\,10$^6$ -- 3\,10$^8$) M$_\odot$ (Tab. 3). 

In order to estimate Eddington ratios, $L_{\rm bol}/L_{\rm Edd}$, where the Eddington luminosity
is given by

\begin{equation}
L_{\rm Edd} = 1.3 \, 10^{38} \frac{M_{\rm BH}}{M_\odot}\,\, \rm{erg\, s^{-1}}, 
\end{equation}

we utilize the close correlation between the luminosity of H$\beta$ and   
the intrinsic continuum luminosity at 5100\AA. For NLS1 galaxies, (Zhou et al. 2006) found that

\begin{equation}
\log(\lambda L_{\lambda}(5100\AA)) = 8.75+0.84 \log {\rm{L(H\beta)}} 
\end{equation}

The bolometric luminosity then follows employing the relation of Kaspi et al. (2000),
$L_{\rm bol} = 9\,\lambda L_{\lambda}$(5100\AA).
All radio-loud NLS1 galaxies of our sample accrete at high Eddington ratios (Tab. 3). 

Alternatively, the MgII-based virial SMBH masses were estimated, following Kong et al. (2006). 

\begin{equation}
M_{\rm BH} = 10^{6.46}\bigg{(}\frac{L_{\rm MgII}}{10^{42}\,{\rm erg\, s^{-1} }}\bigg{)}^{0.57} \bigg{(}\frac{\rm FWHM(H\beta)}{1000\, \rm{km\,s^{-1}}}\bigg{)}^2\, \rm{M_\odot}.     
\end{equation}

These masses agree within better than a factor 2.6 with the H$\beta$-based masses (Tab. 3).  
\subsection{Jet power} 

Radio data of the four NLS1 galaxies were collected from the literature (Tab.1), and are
based on non-simultaneous observations from the FIRST/NVSS survey at 1.4 GHz (Helfand et al. 2015, 
Condon et al. 1998), the GB survey at 4.85 GHz (Gregory \& Condon 1991), 
data at 22 GHz (Doi et al. 2016),
and observations at 151 MHz (Hales et al. 2007). The spectral index $\alpha_{\rm r}$ was calculated
between 1.4 and 4.85 GHz, and defined as in $f_\nu \propto \nu^{\alpha_{\rm r}}$ (Tab. 1). Since 
for SDSSJ1634+4809 no measurement at 4.85 GHz is available, $\alpha_{\rm r}$=0 was assumed. 
The intrinsic radio power $\nu L_\nu$ at rest-frame 1.4 GHz is given by      
 
\begin{equation}
L_{\nu} = 4\pi r_{\rm l}^2 f_{\nu} (1+z)^{-1-\alpha_{\rm r}},  
\end{equation}

where $\nu$=1.4 GHz, $r_{\rm l}$ is the luminosity distance, and $f_{\nu}$ is the observed flux at 1.4 GHz (corresponding
to a higher frequency in the source rest frame).   
We then use this number to perform an order of magnitude estimate for the jet power, 
$P_{\rm jet}$, following Birzan et al. (2008; their equ. 16). Results are listed in Tab. 3. If the
relation of Cavagnolo et al. (2010) is employed 
instead, higher values of $P_{\rm jet}$ are predicted by up to an order of magnitude.    
The jet Eddington ratio is defined as 
\begin{equation}
\eta_{\rm jet}=P_{\rm jet}/L_{\rm Edd}
\end{equation}
and given in Tab. 3.  

\begin{table*}
\centering
\caption{Galaxy properties. Columns from left to right: (1) Galaxy name, (2) bolometric luminosity,  
 (3) H$\beta$-based SMBH mass in solar masses, (4) Eddington ratio,
  (5) MgII-based SMBH mass in solar masses,  (6) jet power in 10$^{44}$ erg\,s$^{-1}$,
  (7) jet Eddington ratio, (8) ionized gas mass in outflow, in 10$^7$M$_\odot$.}
\label{tab:Table3}
   \begin{tabular}{cccccccc}
   \hline
name & log $L_{\rm bol}$ & log $M_{\rm BH,H\beta}$ & $L/L_{\rm Edd}$ & log $M_{\rm BH,MgII}$ & $P_{\rm jet}$ & $\eta_{\rm jet}$ & $M_{\rm out}$  \\
(1) & (2) & (3) & (4) & (5) & (6) & (7) & (8) \\ 
\hline
SDSSJ1305+5116 & 46.5 & 8.4 & 0.95 & 8.1 & 5.0 & 0.015 & 8.7 \\
SDSSJ1443+4725 & 45.6 & 7.6 & 0.79 & 7.9 & 6.1 & 0.11 & 1.6 \\
SDSSJ1505+0326 & 44.7 & 6.8 & 0.66 & 7.0 & 4.3 & 0.52 & 0.3 \\
SDSSJ1634+4809 & 45.4 & 7.5 & 0.65 & 7.1 & 1.4 & 0.03 & 0.3 \\
 \hline
 \end{tabular}
\end{table*}

\section{Notes on individual objects} 

{\bf{SDSSJ1305+5116.}} This galaxy hosts the most massive SMBH in our mini sample (Tab. 3), at the very
upper BH mass range observed for NLS1 galaxies (e.g., Xu et al. 2012). It shows the second highest
outflow velocity in [NeV]. 5 GHz VLBA radio images reveal a core--jet structure at parsec scales, 
and SDSSJ1305+5116 has one of the highest
brightness temperatures of the NLS1 sample of Gu et al. (2015). At kpc scales, a second bright radio
source is detected (Berton et al. 2018). Liao et al. (2015) reported evidence
for a $\gamma$-ray flare from SDSSJ1305+5116.   

{\bf{SDSSJ1443+4725.}} It shows the highest outflow velocity, $\Delta v$([NeV]) = 2450 km s$^{-1}$
and is the strongest FeII emitter, with R4570=2.2, in our sample. While the 1.4--4.85 GHz spectrum, based
on non-simultaneous observations, is steep (Tab. 1), Berton et al. (2018) report a flat spectrum. 
On the other hand, the 5 GHz VLBA
image is resolved into multiple components all of them of steep spectrum (Gu et al. 2015). 
 There is
evidence for faint extended radio emission on kpc scales (Berton et al. 2018). 
A tentative
$\gamma$-ray detection was presented by Liao et al. (2015). 

{\bf{SDSSJ1505+0326.}} It is the brightest radio source of our sample, and the radio loudest, 
with an inverted radio spectrum (Tab. 1), beamed and highly variably at radio 
frequencies (Angelakis et al. 2015, L{\"a}hteenm{\"a}ki et al. 2017). 
It shows a compact core-jet structure on parsec scales
(Dallacasa et al. 1998, D'Ammando et al. 2013), with mildly 
superluminal motion (1.1c$\pm$0.4c; Lister et al. 2016). 
SDSSJ1505+0326 
was detected at $\gamma$-ray energies by the {\it{Fermi}} 
satellite (Abdo et al. 2009b) including recent flaring activity 
(Paliya \& Stalin 2016, D'Ammando et al. 2016). 
Applying BLR scaling relations, its H$\beta$-based virial black hole mass of 6.3\,10$^6$ M$_\odot$ (Tab. 3)   
is relatively
low given its high radio loudness. However, independent estimates show that the BH mass
is still uncertain by a factor of a few, and perhaps more than an order of magnitude. 
Yuan et al. (2008) reported 4\,10$^6$ M$_\odot$,
while Berton et al. (2016) gave 1.8\,10$^7$ M$_\odot$.
Using MgII, 
a virial BH mass of 1.0\,10$^7$ M$_\odot$ was inferred (Tab. 3), 
while estimates from optical--UV SED models of the accretion disc emission range
between 2\,10$^7$ M$_\odot$ (Abdo et al. 2009b), 4.5\,10$^7$ M$_\odot$ (Paliya \& Stalin 2016)
and 2\,10$^8$ M$_\odot$ (Calderone et al. 2013), while D'Ammando et al. (2016) concluded,
that BH masses above few\,10$^7$ M$_\odot$ are inconsistent with the UV SED.  
SDSSJ1505+0326 is mildly variable in the optical (Paliya et al. 2013) and IR (Jiang et al. 2012). 

{\bf{SDSSJ1634+4809.}} Its 5 GHz VLBA image shows only a compact core (Gu et al. 2015). Caccianiga
et al. (2015) estimated a high star formation rate, but concluded, that the radio emission is  
jet dominated. SDSSJ1634+4809 shows the smallest ionized gas velocities of our sample,
with $\Delta v$([NeV]) = 520 km s$^{-1}$.    

\section{Discussion}

\subsection{Mass of outflowing gas} 
 
It is interesting to perform an estimate of the mass of ionized gas which is involved in the outflow. 
As has become common for such type of estimates, it is assumed that 1/10 of 
the [OIII] luminosity
in the outflowing gas is associated with H$\beta$
emission, consistent with estimates in Sect. 2.3. 
The H$\beta$ luminosity is given by 

\begin{equation}
L_{\rm H\beta} = {\int}{\int}j_{\rm H\beta} d\Omega dV, 
\end{equation}
where the gas emissivity 

\begin{equation}
j_{\rm H\beta} = \frac{n^2}{4\pi}\, 1.24\, 10^{-25}\, \rm{erg\,s^{-1}\,cm^{-3}\,ster^{-1}} 
\end{equation}

under case B recombination conditions and at $T$=10.000 K (Osterbrock 1989). 

The mass of the outflowing gas is then given by 

\begin{equation}
M_{\rm out} = 6.74\, 10^7 \bigg{(}\frac{L_{\rm H\beta}}{10^{42}\, \rm{erg\,s^{-1}}}\bigg{)} \bigg{(}\frac{n}{100\, \rm{cm}^{-3}}\bigg{)}^{-1} \,\rm{M_\odot}, 
\end{equation}

where $n$ is the gas density. 
In order to perform an estimate of the ionized gas mass in outflow,
a gas density of $ \log n =2$ was used{\footnote{while the gas in the NLR of AGN
is composed of a range of densities (e.g., Komossa \& Schulz 1997, Villar-Martin et al. 2016), 
a constant gas density of 
$\log n = 2$ is often adopted in similar work (e.g., Liu et al. 2013,
Husemann et al. 2016, Brusa et al. 2015) for an order of magnitude estimate, and we follow
the same approach here}}, 
which then implies  
$M_{\rm out} = 0.9 \times 10^8$ M$_{\odot}$ for SDSSJ1305+5116, and values 
of (0.3--1.6) $\times 10^7$ M$_{\odot}$ for the other galaxies
(Tab. 3). We finally note, that the {\em ionized} gas component comprises only a small
fraction of the total gas mass.

\subsection{The driver of the outflow} 

\subsubsection{Models and constraints from observations} 

The outflow velocities of the high-ionization emission-line gas measured
here are remarkably large, and similar to some of the highest velocities 
detected so far (e.g., Aoki et al. 2005, Holt et al. 2003,
Nesvadba et al. 2011, Zakamska et al. 2016).  
The measured velocities of the high-ionization gas are above the escape
velocity of the host galaxy.

What is the driving force of these flows ?
Radiation pressure acting on dust (e.g., Binette 1998, 
Dopita et al. 2002, Fabian et al. 2006, Thompson et al. 2015) 
has been shown to be capable of accelerating dusty gas to significant velocities.  
Thermal winds need efficient acceleration (and deceleration) mechanisms,
and have been employed to explain mild emission-line 
shifts in nearby Seyfert galaxies (e.g., Das et al. 2005).
High radio luminosity is known to have an influence on the [OIII] line width
(Mullaney et al. 2013). 
Jet-cloud interactions are another possibility to drive 
local gas flows  (e.g., Bicknell et al. 2000, Saxton et
al. 2005, Wagner et al. 2012), including  
the possibility of entrainment of NLR clouds
in the radio plasma of the jets. Clouds are prone to instabilities
under these conditions (e.g., Blandford \& K{\"o}nigl 1979, Schiano et al. 1995), but 
magnetic confinement has been suggested as a way out (Fedorenko et al. 1996).
Accretion disc winds, seen in some cases as ultrafast outflows in X-rays up to $\sim$0.1c
(Tombesi et al. 2012), may proceed into the inner NLR (Proga et al. 2008, Kurosawa \& Proga 2008)
and beyond, and may then shock and accelerate the NLR clouds and drive a large-scale
outflow (Wagner et al. 2013, Cielo et al. 2018). 

Here, we comment on the constraints on the different outflow scenarios,
set by the reported observations. 

\paragraph{Pole-on view} Orientation effects will contribute to increasing
the observed radial velocities, if we have near pole-on views into the outflow cones,
as expected for radio-loud, beamed sources. This will plausibly contribute 
to the higher fraction of blue outliers among radio-loud NLS1 galaxies (Yuan et al. 2008). 
At the same time, among the NLS1s of our mini-sample, the radio-loudest 
and strongly beamed source, SDSSJ1505+0326, does not show the highest outflow velocity, so that
other factors must also play a role.  

\paragraph{Cloud acceleration by radiation pressure} 
Recent simulations have shown that radiation pressure acting on dust can
lead to high outflow velocities. IR photons which undergo multiple scattering events
on dust in dense clouds efficiently transfer momentum to the gas, and this mechanism
is capable of driving large-scale outflows 
with velocities up to 100-1000 km s$^{-1}$ in luminous quasars (Bieri et al. 2017). 
Such a scenario would also explain, why the observed Balmer-line emitting 
BLR (in the NLS1 galaxies of our study) is not involved in the flow, 
since the flow would only be launched beyond the dust-survival
radius.
However, we
do not find strong evidence for dust extinction in the optical spectra.
First, the spectra are very blue. This still leaves the possibility
of dust entrained with the narrow-line emitting gas, of which only a small
fraction partially covers the continuum source. However, second, the observed ratio of
[NeV]/[OIII] is very high, 
whereas strong extinction would significantly decrease [NeV].

\paragraph{Fraction of matter-bounded clouds}  While detailed photoionization modelling
would be needed to understand all emission line intensities, the high flux ratio of [NeV]/[OIII]
stands out. The average intensity ratio $f_{\rm Ne}$ of 0.54 of our mini-sample
exceeds significantly the ratio observed in nearby Seyfert 
galaxies (Komossa \& Schulz 1997) and the average of SDSS quasar 
spectra ($f_{\rm Ne}$=0.16; vanden Berk et al. 2001), and points 
to a higher-than-usual fraction of matter-bounded clouds in
the NLR (e.g., Binette et al. 1997, Komossa \& Schulz 1997); i.e., clouds of low column density with
a higher degree of ionization, and which are more efficiently accelerated by jets or winds (Sec. 4.2.2).  
  
\paragraph{Two-component NLR} 
The [OIII] emission line profile analysis has shown, that there is little 
high-ionization gas at zero velocity. We therefore do not observe a 
two-component NLR in which a significant fraction of the NLR gas is at rest,
while another NLR component is in outflow.  
Rather, a large fraction of the observed NLR participates 
in the outflow{\footnote{At the same time, the bulk of the 
low-ionization H$\beta$-emitting  BLR is not part of
this outflow, implying that either cloud acceleration occurs beyond the BLR
scales, or else that only the high-ionization part of the BLR participates
in the outflow as observed in quasars (review by Marziani et al. 2017).
}}.
The observed ionization stratification points in the same direction. 


\paragraph{Localized jet-cloud interactions of a {\em narrow} jet} 

We have inspected the dependence of radial velocity on
ionization potential for all measured emission lines (Fig. 4).
For comparison, the values measured for our previous sample of blue outliers
are shown (K08), drawn from radio-quiet objects.
The four galaxies follow the same pattern seen before. In particular, the higher
ionization line [NeV] which is also formed at higher critical density, shows
larger outflow velocity than the [OIII] emitting gas. This implies that 
either, the [NeV] emitting gas is more efficiently accelerated than the 
[OIII] emitting gas. Or else, that we see a large-scale outflow, which
{\em decelerates} between the [NeV] emission site further in, and the [OIII] emission 
site further out.    
These observations, and the absence of a second, undisturbed NLR component      
argue against single, localized jet-cloud
interactions.   

\paragraph{Source youth/ early blow-out phase} 
It is interesting to compare the phenomenon of blue outliers in
NLS1 galaxies with that of outflows in radio galaxies, since both
source types are likely in an early evolutionary stage. 
The phenomenon of emission-line shifts is also known among
some young radio galaxies (e.g., Tadhunter et al. 2001,
Marziani et al. 2003, Stockton et al.
2007, Holt et al. 2003, 2008, Nesvadba et al. 2008, 2011),
and is often accompanied by significant outflows in the cold neutral gas
(e.g., Morganti et al. 2010, 2013a, 2013b, Mukherjee et al. 2018). These
galaxies likely represent an early stage of radio-source
evolution, where a joung jet interacts with a cocoon of
dense gas in the central region. 
At least a fraction of these sources is likely
triggered by mergers.
The favored mechanism for producing the line
blueshifts is jet-cloud interactions in the inner NLR, and
the blue outliers among the radio galaxies generally lack
an ionization stratification (Holt et al. 2008), unlike the NLS1 galaxies, suggesting
some difference in the evolutionary state, and/or the scale and driver of the outflows
(we come back to a variant of this scenario in Sect. 4.2.2).   

Since the radio-loud NLS1s of our sample show {\em both}, 
{\em near-Eddington accretion} likely associated with significant {\em outflows}, 
as well as powerful 
{\em radio jets}{\footnote{Since all our sources are very radio-loud, this statement holds
independent of a possible {\em beaming} correction. The estimate of radio power itself (our Sect. 2.10) is known
to be uncertain, and we have therefore used the more conservative estimate (i.e., the one which predicts much lower radio
powers (Sect. 2.10)). Even if the radio powers reported in Tab. 3 
were overestimated by a factor of 10, our salient conclusions
would hold, in that they would still marginally be able to accelerate gas to the high observed velocities, if the NLR clouds
are significantly fragmented (Wagner et al. 2012)}}, we next compare our salient results 
(especially: high outflow velocities, strong [NeV], higher line broadening for
higher velocity shifts, presence
of an ionization or density stratification)
with state-of-the-art simulations of NLR-wide outflows, 
in order to assess under which conditions the observations can be best reproduced.

\subsubsection{Implications from large-scale simulations of the interaction of disc winds and radio jets
with NLR gas} 

Hydrodynamic simulations of jets and winds interacting with the inhomogeneous interstellar medium (ISM) 
of the host galaxy give clues toward interpreting the data presented in this paper. Wagner et al (2011, 2012) 
performed relativistic hydrodynamic simulations of AGN jets interacting with a clumpy ISM on kpc scales, 
while Wagner et al (2013) performed similar simulations for the case of disc winds. 
More recently, 
Bieri et al (2017) performed radiation hydrodynamic simulations of galactic outflows driven mainly 
by the radiation pressure of infrared photons on dust.
 
These were all simulations of isolated galaxies modeled with idealized but realistic initial conditions 
employing a two-phase fractal ISM. The spatial resolution of the simulations was 
approximately 2 pc, allowing gas densities of up to a few 10$^4$ cm$^{-3}$ to be sampled. The 
simulations were designed so that most of the AGN--ISM interactions occur on 1 kpc scales or less. 
The density inhomogeneities, dense cool clouds with temperatures between 30 K and 30000 K, 
in these simulations at small radii can therefore be associated with NLR clouds. The 
key to modeling realistic AGN outflows are the initial conditions of the inhomogeneous ISM. 
In all simulations, powerful multi-phase gas outflows were generated by efficient momentum 
and energy transfer from the jet, or wind. 
The velocity dispersion and radial outflow velocities reached scale with AGN power, and are 
sensitively dependent on the spatial distribution and mean column density of clouds. The higher 
the mean column densities, the harder it is to ablate and accelerate the clouds. 


In the case of jets and winds, the jet or wind plasma percolates through the porous interstellar 
medium generating shear instabilities at the cloud interfaces, ablating and entraining the outer 
layer of clouds, and driving radiative shocks into the interior of the clouds. Clouds at all 
radii experience strong hydrodynamic ablation as well as core compression. The ablated cloud 
material was typically accelerated to 1000s km s$^{-1}$, while bulk clouds gained velocities of 100s 
km s$^{-1}$ (e.g., Wagner et al 2011, their Fig 4). The momentum and energy transfer through jets and winds 
proceeds in a similar fashion, with jets exhibiting, on average, a somewhat larger momentum boost 
(mechanical advantage) and a stronger alignment signature, that is, material along the jet axis 
is more strongly accelerated and compressed than material far away from the jet axis (Wagner et 
al 2016). On the whole, however, feedback by jets and winds are difficult to distinguish in terms 
of the kinematics of the dispersed gas. In fact, the kinematics of the dispersed gas is much more 
sensitive to its structure and spatial distribution at the time the SMBH becomes active.

On kpc scales, AGN radiation on its own is less efficient than jets or fast winds in driving 
outflows (Cielo et al 2018), but momentum boosts of tens of $L_{\rm bol}/c$ can be achieved through 
infrared photons trapped in dusty clouds, which provide most of the momentum boost (Bieri et al 
2017). 
Optical photons provide some momentum, and UV photons ionize and heat the outer layers of 
the clouds, but both are quickly reprocessed into infrared photons. Although shocks are driven 
into the clouds, photons penetrating the clouds rarify the gas within, rendering the clouds 
more diffuse than in the case of compression and acceleration by radio plasma.

The simplest explanation for the ionization stratification or density stratification (Sect. 2.7) 
is that one is observing a spatially correlated outflow that is decelerating with radius. This is 
in line with the trend in observations of nearby spatially resolved AGN that the [NeV] emission 
is more compact than the [OIII] emission. Global deceleration of the outflow is seen in the 
later stages of the simulations with jets or winds, when the jet or wind plasma escapes the 
region covered by clouds and begins to blow a bubble beyond  $\sim$1 kpc (see also Mukherjee et al 2016). 
In the case of radiation driven outflows, the coupling between radiation and gas diminishes 
and the outflow decelerates at later times (after a few Myr), because the efficiency with 
which outflows are accelerated depends primarily on the infrared optical depth, and the 
optical depth rapidly decreases with time as the gas is dispersed. 

The deceleration in all cases is primarily through hydrodynamic drag. However, the deceleration 
seen in the simulations occurs globally and is mainly a function of time, rather than radius or 
density. Even at later times, the clouds at large radii and lower density generally have higher 
velocities than at smaller radii and larger densities.  Note, however, that these simulations 
do not probe AGN--ISM interactions for much longer than a few Myr, and gas may 
couple differently with the jet, wind and radiation when the dispersed clouds fall back in 
again through gravity. Furthermore, the simulations may not be resolving very small, fast-moving 
fragmented clouds at small radii, as the Jeans length is always only marginally resolved in 
these simulations.

Other deceleration mechanisms that are only marginally captured are 1) deceleration by gravity, 
which becomes important on scales exceeding a few kpc (i.e. beyond the baryonic scale radius;
however they are relatively unimportant for those galaxies of our sample with the highest velocities,
well beyond the escape velocity) 
or on scales below a few tens of pc (i.e., the BH sphere of influence); and 2) deceleration 
by backflows. The latter is particularly relevant to jets and can, under certain circumstances, 
even influence gas flows down to parsec scales (Cielo et al 2014).

Another possibility that may contribute to a density stratification is efficient
radiative cooling of diffuse fast filaments of entrained gas. These filaments can exceed
1000 km s$^{-1}$ well within 1 kpc, and in most of the simulations described above, the
cooling time of the filaments is comparable to the dynamical time of the outflow
over ~1 kpc (e.g. Mukherjee et al. 2018). At small radii, ablation may be more efficient
than contraction due to cooling, but, toward large radii, the accelerated filaments
may condense, fragment, and collapse, experiencing reduced hydrodynamic drag and
retaining much of the momentum they gained. The existence of fast dense gas at
large radii may be consistent with the fact that (at least the low-ionization part of)
the BLR does not appear to contribute to the outflow, but it may be at odds with the
compact nature of [NeV] (seen in many spatially resolved AGN, and reflecting the fact
that the emission line requires a high ionization parameter $U \propto L n^{-1} r^{-2}$). 
If, however, this is
the dominant reason for the density stratification, then such a stratification may imply
feedback by jets or winds, rather than by radiation pressure for which radiative shocks
and cooling are likely not as efficient. This effect is not well captured in simulations
thus far, as numerical diffusion often dominates the realistic thermal history of the
accelerated filaments. Global simulations would require higher levels of adaptive mesh
refinement, but the fragmentation due to radiative cooling is seen in wind-tunnel simulations
relevant to galactic outflows by Cooper et al (2008) and Dugan et al (2017).

Finally, a third possibility is that we are observing sources exhibiting ionization or density 
stratifications at early times in the evolution of the AGN outflow. In that case, denser 
clouds at smaller radii have experienced more acceleration than more diffuse clouds at 
larger radii, simply because the AGN bubble or the radiation has not had the time to 
influence the outer clouds much. In this case, however, we might expect a component 
of the NLR to be at rest, which is not seen in our sample -- the entire NLR appears 
to be participating in the outflow. In the case of a jet that is still confined at 
smaller radii, the clouds at larger radii may be accelerated by the forward shock 
sweeping up the NLR and by secondary jet streams that propagate away from the 
main jet stream, percolating through the inter-cloud space all the way to the boundaries 
of the bubble. 
In general, this hypothesis requires that we are observing 
these AGN at a special time within a timeframe of $<$ 1 Myr.  
Overall, this is consistent with other lines of evidence, that NLS1 galaxies
are young AGN in an early stage of their evolution (e.g., Mathur 2000a,b, Grupe 2004),
and with the compactness of their radio emission (e.g., Komossa et al. 2006, 
Gu et al. 2015, Doi et al. 2016). None of the radio-loud galaxies of our sample shows large-scale
radio emission (i.e., they are unrevolved by FIRST; Yuan et al. 2008), and
the bulk of the radio emission arises on scales less than $\sim$200 pc (Gu et al. 2015).  
However, Berton et al. (2018) 
reported evidence for faint extended radio emission on kpc scales in SDSSJ1443+4725, 
SDSSJ1505+0326, and SDSSJ1634+4809, and the
presence of a second bright radio source at 8 kpc separation in SDSSJ1305+5116. 
This is {\em very} compact
when compared to the majority of radio-loud AGN, but is of the dimension
of the NLR. Deeper radio observations at tens of kpc scales of our sample are needed to further constrain
the dimensions of the radio emission at or beyond NLR scales.   

\begin{figure}
\includegraphics[width=\columnwidth]{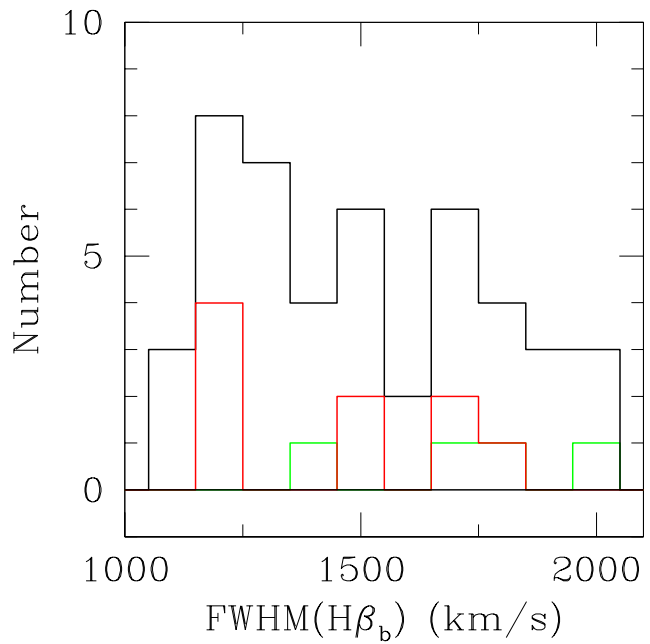}
\caption{Histogram of the FWHM(H$\beta$) distribution of the sample of NLS1 galaxies
of Xu et al. (2007).  The histograms of the blue outliers of Komossa et al. (2008)
and of the radio-loud NLS1 galaxies of this study are plotted for comparison,
in red and green, respectively. Radio-loud NLS1 galaxies do not show
systematically smaller line widths.
 }
 \label{hbetawidth}
 \end{figure}

In a recent comparison study between jet feedback and feedback by radiation pressure, 
Cielo et al (2018) found that the phase space evolution at early times of 
radiation-driven outflows is different to that of jet-driven outflows. In the 
case of radiation-driven outflows, the inner regions of the clouds were 
accelerated first, followed by the more diffuse shadow regions, while in 
the case of jet- or wind-driven outflows, the outer regions of a cloud were 
first quickly ablated and accelerated, and the cloud cores were then accelerated 
in bulk by external pressure gradients. This leads to regions of positive correlation 
between outflow speed and density for up to $\sim$1 Myr in the case of radiation-driven 
outflows, while the correlation is already negative by then in the case of jet-driven 
outflows (Cielo et al 2018, their Fig. 6). The positive correlation in the former case 
disappears at later times.
 
In summary:  
(1) Our results are consistent with several key predictions from the simulations, in that  
the highest outflow velocities are reached in the two galaxies with the highest bolometric
luminosities and jet powers (SDSSJ1305+5116 and SDSSJ1443+4725), there is evidence for a 
population of fragmented, matter-bounded clouds (high [NeV]/[OIII] ratios) 
which are more efficiently accelerated to high velocities, 
and the small implied source lifetimes
are in line with other evidence for the youth of NLS1 galaxies.   
(2) A well-defined ionization or density stratification remains more challenging to understand,
but will provide tight constraints for future models of mechanically driven feedback by jets or winds 
and feedback driven purely by radiation pressure, especially once global radiation-hydrodynamic 
simulations are coupled with non-equilibrium ionization calculations for all the observed species.  

\subsection{A note on orientation effects in NLS1 galaxies} 

While there is now several lines of evidence that the
main driver of NLS1 activity and phenomenology is near-
Eddington accretion onto low-mass black holes (e.g.,
Boroson 2002, Grupe et al. 2010, Xu et al. 2012), 
there is also circumstantial
evidence that orientation plays a secondary role in affecting the
observed properties of NLS1 galaxies (e.g., Collin et al.
2006). Ever since the identification of NLS1 galaxies as AGN
with the smallest widths of the broad Balmer lines, the
question has been around, whether near face-on views onto 
flattened BLRs might contribute to the narrow width of the lines (e.g.,
Osterbrock \& Pogge 1985).  

In case of a heavily flattened BLR, the pole-on NLS1 galaxies
(i.e. the beamed ones, and/or those with the strongest
outflow components), should show a preference for more
narrow Balmer lines. 
We have plotted in Fig. 6 the distribution of
FWHM(H$\beta$) for a sample of 46 NLS1 galaxies (Xu et al. 2007, K08),
excluding blue outliers.
For comparison, the blue
outliers of that sample (K08) are overplotted, as are the four 
NLS1 galaxies
of the present study. 
Note that in order to
make this comparison, we need to rely on alternative 
criteria for NLS1 classification, not using FWHM(H$\beta$).
In fact, all NLS1 under study are still classified as such if we
only use the ratio of [OIII]/H$\beta$ and the strength of
FeII as classification criterion.
We find that the galaxies with the highest outflow velocities do
not show a preference for the smallest Balmer lines, arguing against
a highly flattened BLR in these sources. 
Since the sample size is small, this test has to be repeated with
a larger number of sources.

\section{Summary and conclusions}

The class of NLS1 galaxies is known to show, on average, higher [OIII] outflow velocities
than the class of BLS1 galaxies (e.g., K08). Even larger outflow velocities are found in  
the {\em radio-loud} NLS1 galaxies under study. These are therefore important laboratories 
for understanding large-scale, fast, outflows in AGN.    

We find extreme outflow velocities (up to $\Delta v$ = 2450 km s$^{-1}$) and highly broadened
`narrow' lines (up to FWHM = 2270 km s$^{-1}$)  in four radio-loud NLS1
galaxies in [NeV], and up to $\Delta v$ = 480 km s$^{-1}$ in [OIII].

It is the {\em whole} [OIII] emission-line profiles which
show these kinematic shifts. {\em Additional} blue wings in [OIII] are also
present, and are more highly shifted than the cores. 

The highest-ionization gas is above the escape velocity of the host
galaxy.

The masses of the ionized gas in outflow are on the order of (0.3--8.7)\,10$^7$M$_\odot$.  

The emission line width shows a very strong correlation with outflow velocity,
so that the highest-velocity gas component inevitably has very 
high velocity dispersion.

The presence of an ionization stratification 
(i.e., a correlation of outflow velocity with ionization potential), and the absence
of significant zero-velocity high-ionization gas,  disfavors single, localized jet-cloud
interactions as the cause of the outflow. Rather, a large fraction
of the NLR must be in outflow. 

Comparison with hydrodynamic models of jets and winds interacting at NLR scales suggests that
we may see NLS1s with high ionized gas velocities in an early stage of their
evolution, in a scenario where the denser, high-ionization clouds at smaller radii have experienced
more acceleration than more diffuse clouds further out. In this scenario, the outer NLR
has not yet been much affected by the jet per se,  but the clouds may still be accelerated
by the forward shock sweeping up the outer NLR, and by jet streams which propagate away from the
main jet stream and percolate through the inter-cloud funnels. This scenario requires
source life times of typically $<$ 1 Myr.  

In addition,  radio selection (especially beamed sources) might contribute to
the high observed radial outflow velocities, since effects of
polar outflows are more pronounced, if we have a near pole-on view into the central engine. 

In summary, radio-loud Seyfert galaxies with extreme outflows are important test beds for
driving mechanisms of large-scale, high-velocity outflows, for feedback
processes in the relatively nearby universe, and perhaps NLS1 orientation models.

\section*{Acknowledgements}
DX acknowledges the support of the Chinese
National Science Foundation (NSFC) under grant NSFC-11273027.
AYW has been supported in part by ERC Project No. 267117 (DARK) 
hosted by Universit\'e Pierre et Marie 
Curie (UPMC) -- Paris 6, PI J.~Silk, and by the Australian Research 
Council through the Discovery Project, 
The Key Role of Black Holes in Galaxy Evolution, DP140103341.
This research has made use
of the SDSS database, and of the NASA/IPAC Extragalactic Database (NED) 
which is operated by the Jet Propulsion Laboratory, California Institute of
Technology, under contract with the National Aeronautics and Space Administration. 
Funding for the SDSS and SDSS-II has been provided by the Alfred P. Sloan Foundation, 
the Participating Institutions, the National Science Foundation, 
the U.S. Department of Energy, the National Aeronautics and Space Administration, 
the Japanese Monbukagakusho, the Max Planck Society, and the Higher Education 
Funding Council for England. The SDSS Web Site is http://www.sdss.org/.
The SDSS is managed by the Astrophysical Research Consortium for the 
Participating Institutions. 
The Participating Institutions are the American Museum of Natural History, 
Astrophysical Institute Potsdam, 
University of Basel, University of Cambridge, Case Western Reserve University, 
University of Chicago, Drexel University, Fermilab, the Institute for Advanced Study, 
the Japan Participation Group, Johns Hopkins University, the Joint Institute for Nuclear 
Astrophysics, the Kavli Institute for Particle Astrophysics and Cosmology, the Korean 
Scientist Group, the Chinese Academy of Sciences (LAMOST), Los Alamos National Laboratory, 
the Max-Planck-Institute for Astronomy (MPIA), the Max-Planck-Institute for Astrophysics (MPA), 
New Mexico State University, Ohio State University, University of Pittsburgh, 
University of Portsmouth, Princeton University, the United States Naval Observatory, 
and the University of Washington.






%
%


\bsp	
\label{lastpage}
\end{document}